\documentclass[]{aastex}

\usepackage{emulateapj5}
\usepackage{onecolfloat}
\usepackage{graphicx} 
\usepackage{fancyheadings} 
\usepackage{ulem}
\usepackage{rotating}
\usepackage{lscape}

\newcommand{\lya}{Ly$\alpha$ }

\newcommand{\e}[1]{{\epsilon({\rm #1})}}

\newcommand{\rAA}{{\AA \enskip}}

\newcommand{\mkms}{{\rm \; km\;s^{-1}}}

\newcommand{\tskip}{\tablevspace{3pt}}

\begin{document}

\twocolumn[%

\title{ON THE PERILS OF HYPERFINE SPLITTING: A reanalysis of Mn and Sc
Abundance Trends}

\author{ JASON X. PROCHASKA \& ANDREW McWILLIAM}
\affil{The Observatories of the Carnegie Institute of Washington}
\affil{813 Santa Barbara St. \\
Pasadena, CA 91101}

\begin{abstract} 
We investigate the impact of hyperfine splitting on stellar abundance analyses 
of Mn and Sc, and find that incorrect hfs treatment can lead to spurious 
abundance trends with metallicity.  We estimate corrections to a recent study 
by Nissen et al. (2000), and find: 
(1) $\lbrack$Mn/Fe$\rbrack$ is described by a
bimodal distribution, 
with $\lbrack$Mn/Fe$\rbrack$~$\sim -$0.3 for stars 
$\lbrack$Fe/H$\rbrack$~$< -0.7$,
and $\lbrack$Mn/Fe$\rbrack$~$\sim -$0.05 for stars at 
higher metallicity, suggestive of a
transition between halo/thick disk and thin disk populations. 
(2) The large majority of stars show nearly solar $\lbrack$Sc/Fe$\rbrack$ 
ratios; although important deviations cannot be ruled out.
\end{abstract}

]

\pagestyle{fancyplain}
\lhead[\fancyplain{}{\thepage}]{\fancyplain{}{PROCHASKA \& McWILLIAM}}
\rhead[\fancyplain{}{ON THE PERILS OF HYPERFINE SPLITTING}]{\fancyplain{}{\thepage}}
\setlength{\headrulewidth=0pt}
\cfoot{}

\section{INTRODUCTION}

A fundamental goal of stellar abundance studies is to investigate
and identify the nucleosynthetic origin and evolution of individual
elements in various stellar populations. 
Observed abundance ratios may be used to constrain theoretical 
nucleosynthesis yields \citep[e.g.][]{ww95}, and gain insight into
the formation and chemical enrichment history of galactic systems.
\cite{tinsley79} developed a paradigm of chemical evolution to explain
early observations \citep{waller62,conti67} which showed
that oxygen and other 
$\alpha$-elements (e.g. Mg, Si, Ca) are enhanced in metal-poor halo stars
relative to iron, but not in solar-metallicity disk stars
(for more discussion see McWilliam 1997).  
Based on this simple idea one expects certain trends of [$\alpha$/Fe]
with [Fe/H] for systems with different formation time scales 
(i.e. SFR, IMF, and binary fractions); thus the measurement of [$\alpha$/Fe] 
can be
used to probe these parameters, and our understanding of chemical 
evolution.  This also suggests that additional diagnostics of the 
Galactic star 
formation history may be obtained by careful abundance studies of other
elements.
In the following, we consider the abundance trends of Sc and Mn, two so-called
iron-peak elements.  

\cite{waller62} first noted that Mn is 
deficient relative to Fe in low metallicity stars, consistent with
the odd-even effect for iron-peak elements suggested by \cite{helfer59}
and \cite{arnett71}.   
These early observations were later confirmed by \cite{gratton89}, who found
that Mn behaves in an opposite sense to the $\alpha$-elements: a steady decline
to [Mn/Fe]~$\sim -0.34$ from solar metallicity to [Fe/H]~$\sim -1.0$~dex,
followed by a plateau at [Mn/Fe]~$\sim -0.34$ between [Fe/H] = $-$1.0 to
$-$2.4~dex.  Gratton concluded that Mn is under-produced in
early massive stars, and over-produced in type~Ia SN.
Below [Fe/H]~$\sim -2.5$, \cite{mpss2} showed that the [Mn/Fe] ratio
declines rapidly, and suggests the presence of a distinct early population~II,
or population~III in the Galactic halo.
More recently, Nissen et al.\ (2000; N00) published a comprehensive study
of $\approx$120 [Mn/Fe] measurements from F and G dwarf stars
in the metallicity interval $-1.4<[{\rm Fe/H}]<+0.1$ .  Their results found 
the same gradual decline in [Mn/Fe] as \cite{gratton89}, but to lower
[Mn/Fe] values and with the suggestion of a drop in [Mn/Fe] at 
[Fe/H]~$\sim -0.7$ dex.

If Mn is a type~Ia SN product, we expect sub-solar [Mn/Fe] in the
Galactic bulge and giant elliptical galaxies, which are enhanced in
type~II SN products \citep[e.g.][]{worthey92,mr94}.  For the LMC and
dwarf galaxies excesses of elements produced by type~Ia SN are anticipated
\citep{wg91}.
The behavior and origin of Mn is also relevant to interpretation of the 
abundance patterns in damped \lya systems \citep{lu96,pro97,ptt00}.
In nearly every damped system where Mn is
observed, one finds an underabundance of Mn relative to Fe
which suggests an underlying type~II SN enrichment pattern independent of
dust depletion.  This interpretation, however, hinges
on the fact that Mn is primarily produced in type~Ia SN.
Identification of the formation processes for Mn, therefore,
has direct bearing on the nucleosynthetic history of these protogalaxies.

Nucleosynthesis calculations \citep[e.g.][]{tww95} predict
deficient [Sc/Fe] ratios 
in metal-poor stars. This contrasts
starkly with the observations of Zhao \& Magain (1990, 1991) who found
$+$0.3 dex Sc enhancements.  However, \cite{gs91},
\cite{peterson90}, and \cite{mpss2} found no evidence
for a deviation from [Sc/Fe]=0.0, and \cite{gilroy88}
found Sc deficient by $\approx$0.2 dex.  The recent analysis of N00 
supports the enhancements found by Zhao \& Magain, with the surprising
conclusion that Sc behaves like an $\alpha$-element; particularly convincing
is the fact that N00 find relatively deficient Sc in their sub-sample of
$\alpha$-element poor stars.  

\begin{table*}[ht]
\tabletypesize{\scriptsize}
\caption{\centerline
{\sc HYPERFINE SPLITTING OF Mn~I AND Sc~II LINES\tablenotemark{a}}}
\begin{center}
\begin{tabular}{cc|cc|cc|cc|cc|cc|cc}
\tableline
\tableline \tskip
\multicolumn{2}{c|}{Mn~I $\lambda$6013} & 
\multicolumn{2}{|c|}{Mn~I $\lambda$6016} &
\multicolumn{2}{|c|}{Mn~I $\lambda$6021} &
\multicolumn{2}{|c|}{Sc~II $\lambda$5526} &
\multicolumn{2}{|c|}{Sc~II $\lambda$5657} &
\multicolumn{2}{|c|}{Sc~II $\lambda$6245} &
\multicolumn{2}{|c}{Sc~II $\lambda$6604}  \\
\multicolumn{1}{c}{$\Delta \lambda$\tablenotemark{b}} & 
$\log gf$\tablenotemark{c} &
\multicolumn{1}{c}{$\Delta \lambda$} & $\log gf$ &
\multicolumn{1}{c}{$\Delta \lambda$} & $\log gf$ &
\multicolumn{1}{c}{$\Delta \lambda$} & $\log gf$ &
\multicolumn{1}{c}{$\Delta \lambda$} & $\log gf$ &
\multicolumn{1}{c}{$\Delta \lambda$} & $\log gf$ &
\multicolumn{1}{c}{$\Delta \lambda$} & $\log gf$ \\
\tableline \tskip
.478 & $-$0.766 & .619 &  $-$1.460 & .746 & $-$2.668 & .770 & $-$3.051& 
.886 & $-$1.229 & .621 &  $-$1.736   & .582 & $-$2.505   \\
.499 & $-$0.978 & .645 &  $-$1.297 & .772 & $-$1.451 & .775 & $-$2.648&
.888 & $-$1.799 & .629 &  $-$2.476   & .590 & $-$2.347   \\
.518 & $-$1.251 & .647 &  $-$0.682 & .774 & $-$2.316 & .779 & $-$2.436&
.893 & $-$1.799 & .631 &  $-$1.907   & .594 & $-$1.936   \\
.527 & $-$1.455 & .667 &  $-$1.292 & .795 & $-$1.275 & .779 & $-$1.838&
.894 & $-$1.627 & .636 &  $-$3.476   & .596 & $-$2.358   \\
.533 & $-$1.661 & .668 &  $-$0.945 & .798 & $-$2.191 & .783 & $-$2.326&
.895 & $-$1.641 & .638 &  $-$2.293   & .599 & $-$2.333   \\
.538 & $-$1.309 & .684 &  $-$1.276 & .804 & $-$0.533 & .783 & $-$1.633&
.901 & $-$2.323 & .640 &  $-$2.114   & .602 & $-$2.531   \\
.547 & $-$1.330 & .685 &  $-$1.394 & .813 & $-$1.249 & .786 & $-$1.567&
.902 & $-$1.652 & .644 &  $-$3.058   & .604 & $-$3.029   \\
.552 & $-$1.485 & .696 &  $-$1.723 & .817 & $-$2.271 & .787 & $-$2.305&
.904 & $-$1.652 & .646 &  $-$2.260   & .607 & $-$4.455   \\
.562 & $-$1.807 & .696 &  $-$1.460 & .821 & $-$0.673 & .788 & $-$1.571&
.906 & $-$1.825 & .647 &  $-$2.385   & .609 & $-$2.707   \\
.566 & $-$1.853 & .698 &  $-$1.656 & .827 & $-$1.316 & .789 & $-$0.909&
.906 & $-$3.749 & .650 &  $-$2.824   & .611 & $-$2.505   \\
.566 & $-$2.409 & .704 &  $-$2.422 & .834 & $-$0.831 & .790 & $-$2.414&
.908 & $-$1.825 & .651 &  $-$2.319   & .613 & $-$2.347   \\
.567 & $-$2.029 & .707 &  $-$1.297 & .837 & $-$1.492 & .790 & $-$1.632&
.909 & $-$2.001 & .652 &  $-$2.803   & .615 & $-$2.531   \\
     &          & .713 &  $-$1.292 & .843 & $-$1.015 & .791 & $-$1.030& 
     &          & .655 &  $-$2.678   & .615 & $-$2.358   \\
     &          & .714 &  $-$1.656 & .846 & $-$1.522 & .792 & $-$1.754& 
     &          & .656 &  $-$2.502   &      &             \\
     &          & .716 &  $-$1.394 & .847 & $-$1.237 & .793 & $-$1.937&
     &          & .657 &  $-$2.581   &      &             \\
     &          &      &           &      &          & .793 & $-$1.166& 
     &          &      &             &      &             \\
     &          &      &           &      &          & .794 & $-$1.322& 
     &          &      &             &      &             \\
     &          &      &           &      &          & .795 & $-$2.083& 
     &          &      &             &      &             \\
     &          &      &           &      &          & .795 & $-$1.741& 
     &          &      &             &      &             \\
     &          &      &           &      &          & .795 & $-$1.507& 
     &          &      &             &      &             \\
\tableline
\end{tabular}
\end{center}
\tablenotetext{a}{Line data taken from http://cfaku5.harvard.edu}
\tablenotetext{b}{The correct wavelength equals $\lambda + \Delta \lambda$}
\tablenotetext{c}{The absolute $gf$ values have no bearing on our analysis}
\end{table*}

While one of us (JXP) was pursuing an analysis of Mn for a sample of 
thick disk stars \citep{pro00}, we became concerned about the N00 Mn analysis
due to their use of the incorrect hyperfine splittings of 
Steffen (1985; S85).
In particular, we were concerned that an incorrect hfs treatment and the
change in line-strength with metallicity could
lead to an artificial trend of [Mn/Fe] versus [Fe/H].  In the course of
this work we also worried about the N00 treatment of Sc~II lines; 
the large S85 hfs splittings seem to have caused N00
to derive large hfs abundance corrections in contrast to previous work
\citep{mr94,mpss2} which found only small hfs corrections for scandium.
In this paper, we estimate the error in derived Mn and Sc abundances due
to use of the S85 hfs lists and consider the effects of the
corrections on the abundance trends found by N00.  The focus of
this paper is to warn the reader of the uncertainties associated with 
hyperfine splitting.  While we make an estimate of the true Mn and Sc
abundances from the N00 sample, our analysis is limited by the fact
that neither the microturbulent velocities or the iron equivalent
widths were available to us.

\section{HFS ANALYSIS}
\label{sec-hfs}

Similar to other odd-Z elements (e.g.\ V, Co, Cu), 
hyperfine interactions between nuclear and electronic wavefunctions
split the absorption lines of Mn and Sc into multiple components. 
For the majority of the Mn~I and Sc~II lines, the hyperfine components
have typical splittings of up to a few tens of m\AA.  In terms of
stellar abundance studies, the hyperfine splitting 
de-saturates strong absorption lines resulting in features with 
larger equivalent widths than single lines with no hfs components.
Thus, if the splittings are ignored, or under-estimated, the
computed abundances will be over-estimated for strong lines.  
Because the size of the abundance over-estimate increases with line-strength,
it is important to treat hfs effects correctly for strong lines;
otherwise spurious abundance trends may result.
For weak lines the hfs treatment is not so critical because they are
already unsaturated.
To derive an abundance from a line affected by hfs one must perform
a spectrum synthesis over the hfs components; thus 
the number of components, the wavelength splittings, and the relative
strengths must be known.  The number of components and their relative
strengths are readily computed with the equations in \cite{condon35}.
To compute wavelengths it is necessary to have splitting
constants ($A$ and $B$) for each electronic level, preferably from
laboratory measurements.

In this analysis we adopt the hfs line lists of \cite{kur99}.
For the Mn~I lines of N00 the hfs constants are from \cite{hsw69}, based
on laboratory measurements, and
for the Sc~II lines the measured hfs constants are from \cite{mansour89}.  
Typical 1$\sigma$ uncertainties for measured Sc~II A and B constants are less
than 0.2\% and 0.5\% respectively.  For the 5526\AA\  line the uncertainty in
hfs component wavelengths are nearly all less than 0.001\AA , with a
maximum uncertainty of 0.003\rAA for the weakest components; these values are
much smaller than the velocity widths ($\sim$ 0.02\AA ),
and far too small to significantly affect the computed abundance.
The Mn~I lines have rms errors $< 0.5\%$ in the A constants but larger
uncertainties in the B constants.  As with the Sc~II lines, these 
uncertainties are insignificant.

\begin{figure*}[ht]
\begin{center}
\includegraphics[height=3.6in, width=3.3in, angle=-90]{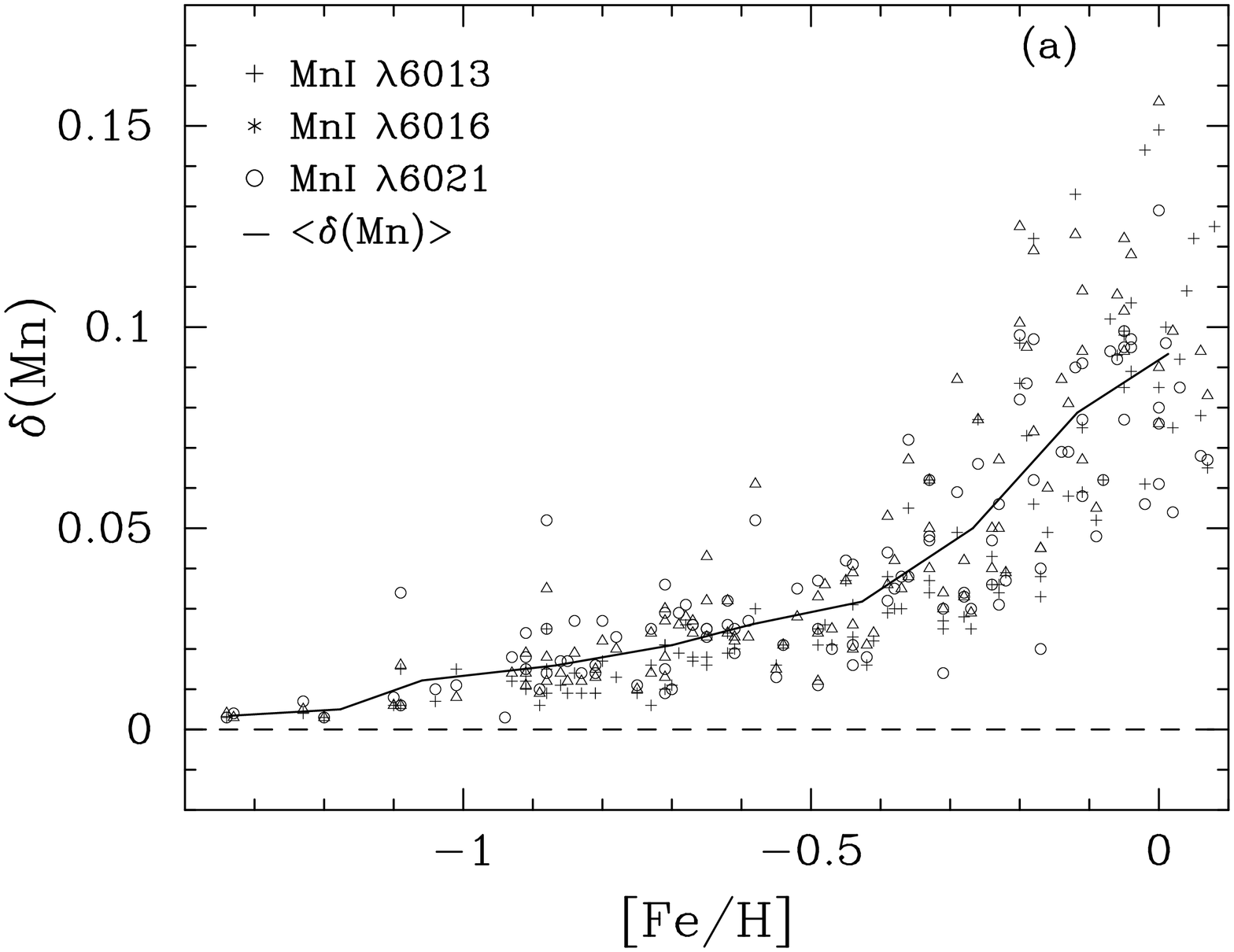}
\includegraphics[height=3.6in, width=3.3in, angle=-90]{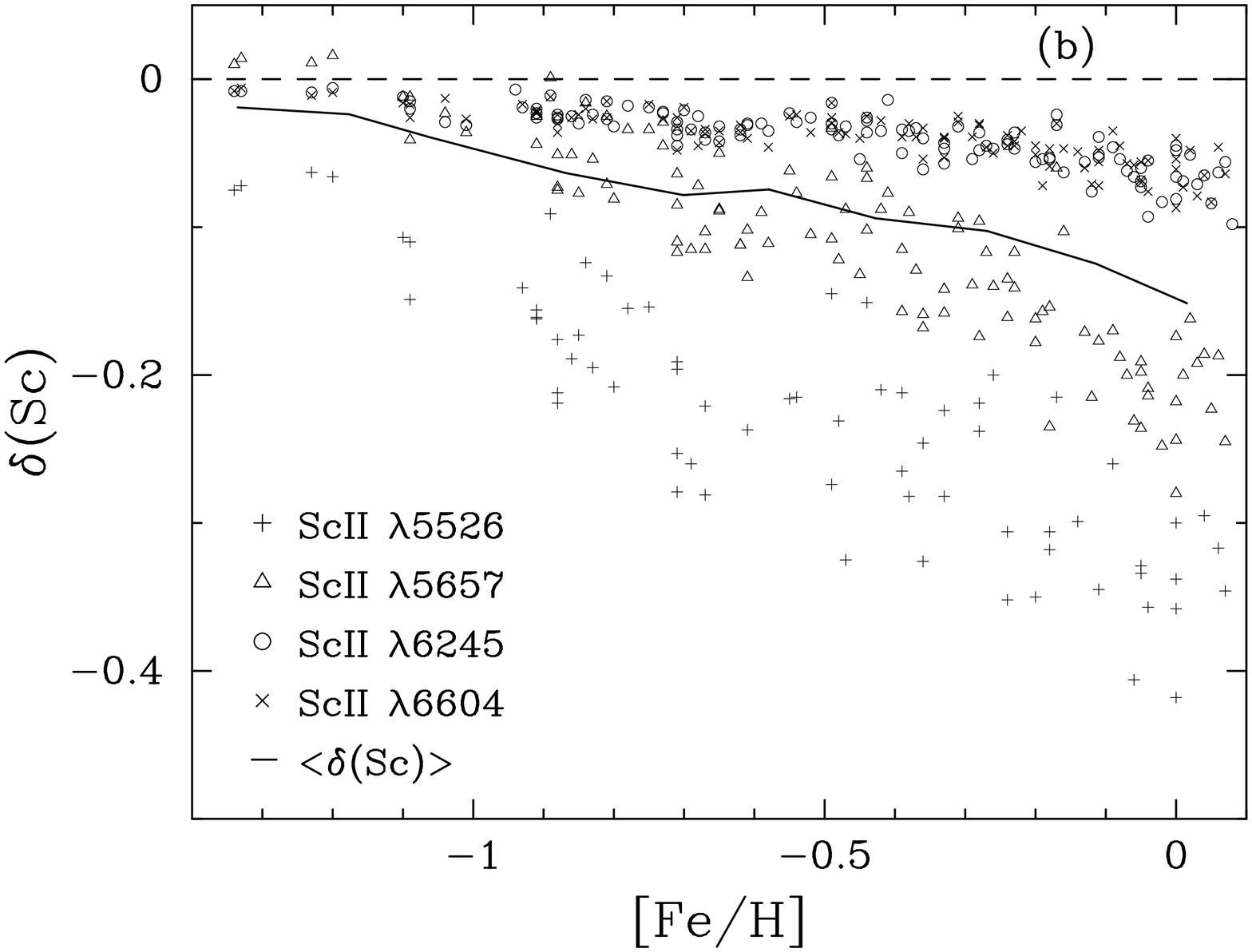}
\caption{Corrections to the Steffen (1985) hfs analysis of (a) Mn~I 
and (b) Sc~II when the more
accurate hyperfine lines listed in Table~1 are considered.  Note that the
corrections have a dependence with metallicity such that ignoring them
might lead to erroneous trends in the Mn and Sc abundances.  In this
analysis, we have adopted the microturbulence relation
from Nissen \& Schuster (1997) with $\xi = 1.15 \mkms$ for the Sun. } 
\label{fig-diff}
\end{center}
\end{figure*}

In a few instances we have checked the \cite{kur99} hfs line parameters.
Table~1  lists the
hfs components for the three Mn~I absorption lines and four of the five
Sc~II lines analyzed in N00.  Unfortunately there are 
no $A$ or $B$ constants measured for the Sc~II $\lambda5239$ line and we
do not consider it in this analysis.
Regrettably, N00 and \cite{gratton89} did not use the best hfs lists.
\cite{gratton89} relied upon the incomplete hfs lists of \cite{booth83}.
However, our analysis of the solar Mn~I lines shows that use of the 
\cite{booth83} hfs list results in a negligible mean error of 
$\approx$0.002 dex. N00 adopted the hfs components published by S85 based
on the work of \cite{biehl76}; 
unfortunately, S85 grouped together nearby hfs components and applied
old, inaccurate, splitting constants.
In the case of Mn~I lines, the true hfs splittings are 
larger than indicated by S85 while for Sc~II lines
the true splittings are much smaller than indicated by Steffen.

In this analysis we 
investigate the differences between the S85 compilation and
the data presented in Table~1 for the N00 sample.  
Because N00 did not publish iron line equivalent widths or 
microturbulent velocities, and because of the different model atmospheres
and synthesis codes employed we could only calculate the abundance 
{\em corrections}
arising from the use of the different hfs line lists.
To compute abundances we employed the \cite{kur93} grid of model
atmospheres, interpolated to the stellar parameters ($T_{eff}$, 
$\log g$, and [Fe/H]) given by N00 and the synthesis program MOOG
of \citep{sneden73}.  Microturbulent velocities for all stars were adopted
based on the relation of \cite{nissen97}.  Use of the 
N00 solar microturbulent velocity, at 1.44 Km/s, would have decreased 
our corrected [Mn/Fe] data points by 0.06 dex.
The Unsold damping approximation, without enhancement, was assumed throughout.

Figure~\ref{fig-diff}a plots the hfs difference,
$\delta({\rm Mn}) \equiv \e{Mn}|_{Nissen} - \e{Mn}|_{PM}$, for every 
measured line in N00 as a function of [Fe/H].
The figure clearly demonstrates that the 
S85 analysis overestimates the Mn abundance and, most importantly,
that the effect is largest at high metallicity.  In turn, the error
leads to an observed trend of decreasing [Mn/Fe] with decreasing metallicity
as described by N00.  Because we compute differential corrections to
the N00 abundances we expect that our results are largely insensitive to errors
in the model atmosphere or the exact assumptions of the spectrum synthesis
code.  While a significantly higher damping or microturbulence parameter 
lessens the effect described by Figure~\ref{fig-diff}a, qualitatively the 
picture is unchanged.  
Analogous to the Mn treatment, we plot
$\delta({\rm Sc})$ vs.\ [Fe/H] in Figure~\ref{fig-diff}b. Here, the 
S85 hfs compilation leads to a significant overestimate of 
the hfs correction, particularly for $\lambda 5526, 5657$.  In turn,
this difference could account for the observed enhancement of Sc described
by N00.

\begin{figure*}[ht]
\begin{center}
\includegraphics[height=3.6in, width=3.3in, angle=-90]{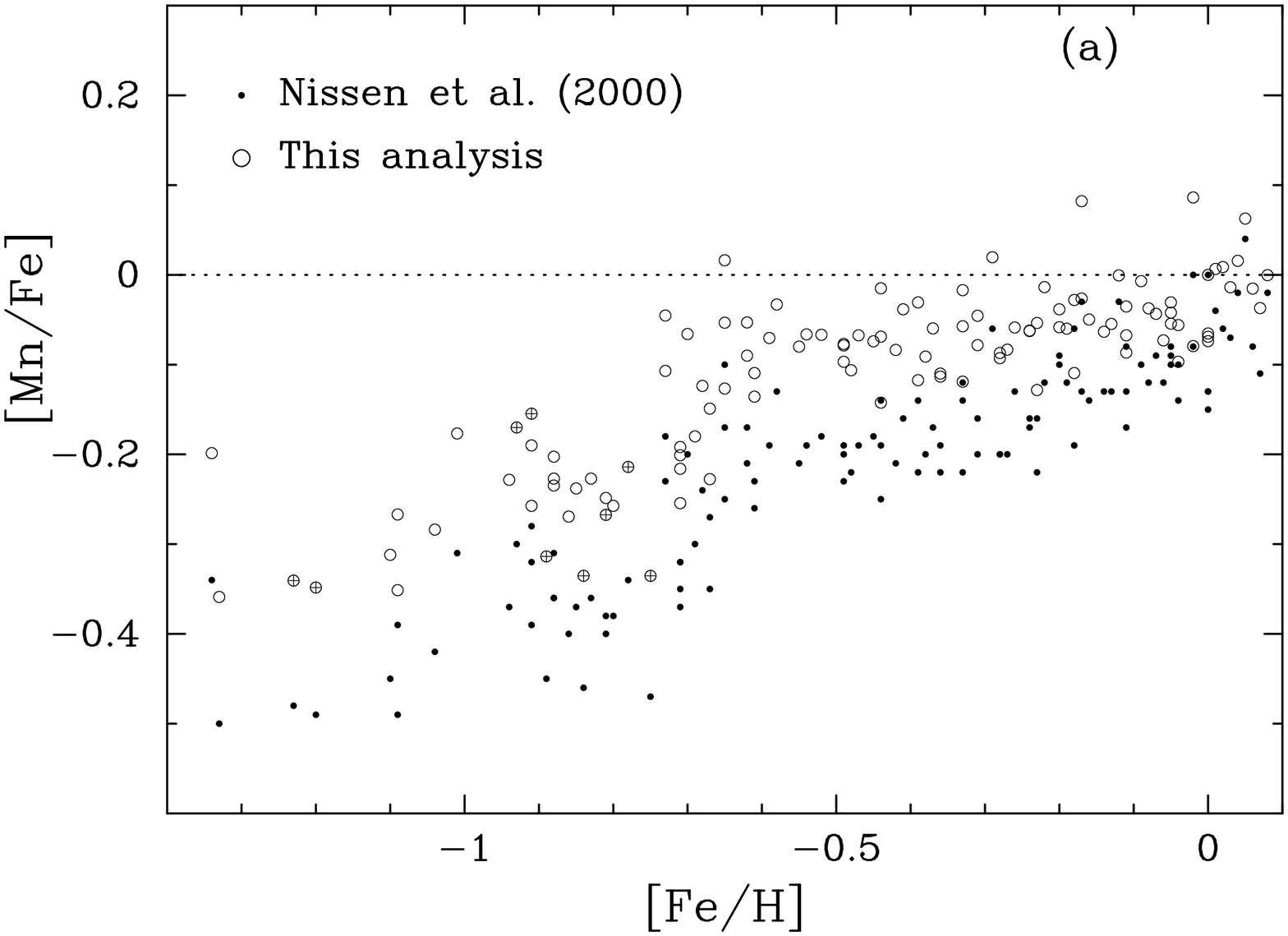}
\includegraphics[height=3.6in, width=3.3in, angle=-90]{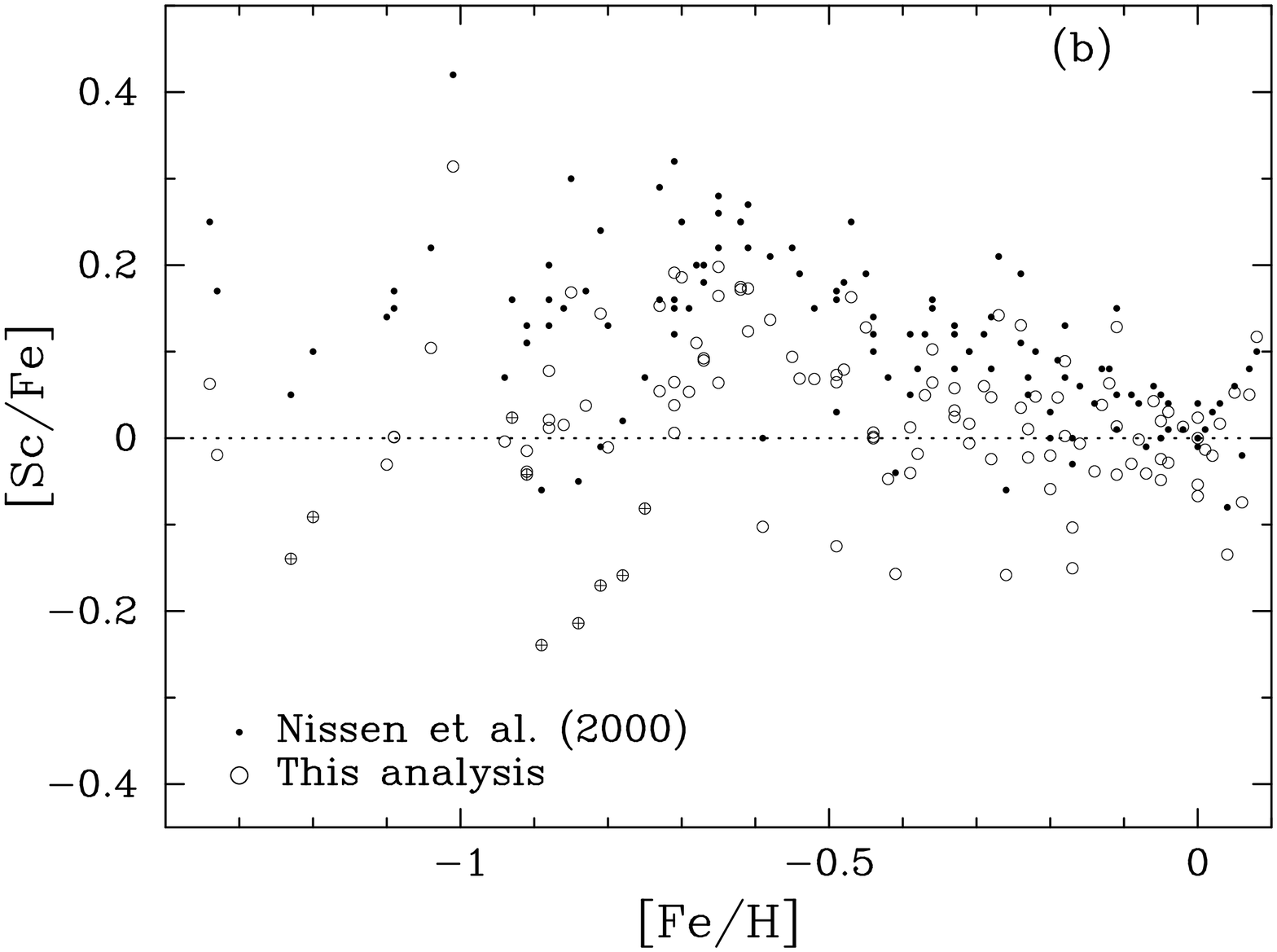}
\caption{A comparison of the Nissen et al.\ (2000) compilation and our
corrected values for the
(a) [Mn/Fe] and (b) [Sc/Fe] abundance ratios vs.\ [Fe/H].  For our analysis,
we have applied line-by-line corrections to the Nissen et al.\ abundances
as described in Figure~1.  The results show that the trends observed in 
the Nissen et al.\ values are quite possibly the result of errors in the 
Steffen (1985) hfs compilation. The +-signs mark $\alpha$-poor stars
identified by Nissen \& Schuster (1997).}
\label{fig-new}
\end{center}
\end{figure*}

\section{DISCUSSION}
\label{sec-discuss}

We now apply the corrections derived in the previous section 
in a line-by-line fashion to the N00 compilation and thereby estimate the
true [Mn/Fe] and [Sc/Fe] trends.  Although our results should qualitatively
describe the metallicity dependence of Mn and Sc, a 
quantitative discussion will await a full reanalysis by the authors
of N00.

Figure~\ref{fig-new}a compares the
corrected [Mn/Fe] values based on our hfs calculations 
versus the N00 results.  
The corrected analysis offer a considerably different picture
from that of N00.  The most robust change is the systematic increase
of $\approx 0.15$~dex in [Mn/Fe] for the most metal-poor stars.  This
[Mn/Fe] shift arises from the substantial hfs corrections to the solar
Mn abundance.  One also notes that the trend toward lower
[Mn/Fe] values from [Fe/H]~=~$0 \to -0.5$ is no longer evident in the
reanalysis.  In short the corrected
[Mn/Fe] values are bimodal; those stars with [Fe/H]~$< -0.7$ have 
[Mn/Fe]~$\sim -0.3$ in agreement with \cite{gratton89} and
the higher metallicity stars show [Mn/Fe]~$\sim -0.05$.  
The most straightforward explanation for these results is
that [Fe/H]~$\sim -0.7$ marks
the passing from the thin disk stellar population to the thick disk and
halo populations. 
The Mn/Fe trends are consistent with Mn being produced primarily by 
type~Ia SN and do not require a metallicity dependent yield.
We note in passing that the plateau 
of Mn/Fe for the thin disk stars lies rather uncomfortably 
below solar: [Mn/Fe]~$\sim -0.05$~dex.  
While the Sun might have a high Mn abundance relative
to the bulk of the thin disk dwarf stars, the offset could be due
to a zero-point error in the N00 solar analysis.

Now consider the results for Sc presented in Figure~\ref{fig-new}b.
In short, the corrected values no longer show the $\alpha$-enhanced trend 
discussed in the N00 analysis. The solar Sc abundance has been
revised upward by $\approx 0.2$~dex which, in turn, has eliminated
the relative enhancement in the metal-poor stars.  
Our results agree well with the Sc abundance analysis performed by
\cite{gs91} indicating that essentially all stars exhibit solar Sc/Fe.
There are two sets of stars, however, with reasonably significant departures
from [Sc/Fe]~=~0: (i) the $\alpha$-poor stars (marked with a +) 
identified by \cite{nissen97} exhibit sub-solar Sc/Fe, consistent with an 
origin related to the $\alpha$-elements; 
(ii) nearly every star with [Fe/H]~$\sim -0.6$ shows enhanced Sc/Fe.  
This trend is difficult to explain and we wonder if it is simply a statistical
anomaly or perhaps a small error in the original analysis of N00.
Finally, we note the two 
weakest Sc~II lines ($\lambda 6245,6604$) give systematically higher
$\e{Sc}$ values than the two stronger lines, particularly in the 
metal-poor stars.
This difference may be explained by blends between the
strong lines of the Solar spectrum and unidentified neutral
metal lines which would lead to erroneously high $gf$ values. 
To account for this systematic offset, we recommend
that future studies on Sc include many more Sc~II lines.

In this Letter, we have investigated the effects of miscalculating the
hfs correction on elemental abundance trends.  We have performed a simple
exercise to correct for the erroneous hfs compilation used by N00 in
their analysis of Mn and Sc.  In turn, we have demonstrated that hfs errors
can mimic metallicity-dependent abundance trends.  Our results highlight
the importance of accurately addressing hyperfine splitting and we encourage
the reader to utilize the large database of hfs lines being compiled by
R. Kurucz.  Having estimated corrections to the Mn and Sc abundances from N00, 
we find the following:  (1) the [Mn/Fe] values are best described by a 
bimodal distribution with stars at [Fe/H]~$< -0.7$ exhibiting 
[Mn/Fe]~$\sim -0.3$ and stars at higher metallicity showing
[Mn/Fe]~$\sim -0.05$.  These trends are consistent with type~Ia SN
being the principal site for Mn nucleosynthesis;
(2) the large majority of stars have nearly solar Sc/Fe ratios consistent
with the results of \cite{gs91} and \cite{mpss2}.  There are exceptions,
however, namely the $\alpha$-poor stars of the N00 sample which show
systematically low Sc/Fe values and the bulk of stars at [Fe/H]~$\sim -0.6$
which exhibit enhanced Sc.

\acknowledgments

We thank R. Kurucz for providing his hfs line lists in a readily
accessible format and for the use of his codes and databases.
We would also like to thank R. Bernstein, B. Weiner, and P. Nissen
for helpful comments. J.X.P. acknowledges support from a
Carnegie postdoctoral fellowship.  A.M. gratefully acknowledges support 
from NSF grant AST-96-18623.

\clearpage

\end{document}